\documentclass[twocolumn,preprintnumbers,amsmath,amssymb,nofootinbib]{revtex4-1}

\pdfoutput=1

\usepackage{graphicx}
\usepackage{dcolumn}

\def\nn{\nonumber}
\def\bea{\begin{eqnarray}}
\def\eea{\end{eqnarray}}
\def\obar{\overline}
\newcommand{\eq}[1]{(\ref{#1})}

\def\a{\alpha}          
           
  \def\C{\Gamma}  
  \def\D{\Delta}

\def\g{\gamma}

\def\s{\sigma}

\def\cA{{\cal A}}  
  
 \def\cH{{\cal H}} 
  
\def\cM{{\cal M}}

\def\R{{\mathbb R}}
\def\C{{\mathbb C}}

\def\one{\mbox{1 \kern-.59em {\rm l}}}

\def\({\left(}
\def\){\right)}
\def\diag{\mbox{diag}}
\def\Tr{{\rm Tr}}

\def\msu{\mathfrak{su}}
\def\mso{\mathfrak{so}}

\begin{document}

\preprint{UWTHPh-2017-39}

\title{Quantized open FRW cosmology from Yang-Mills matrix models }

\author{Harold C. Steinacker}%
 \email{harold.steinacker@univie.ac.at}
\affiliation{Faculty of Physics, University of Vienna\\
Boltzmanngasse 5, A-1090 Vienna, Austria } 

\date{\today}

\begin{abstract}

We present simple solutions of IKKT-type matrix models describing a quantized
homogeneous and isotropic cosmology with $k=-1$,  finite density of microstates and a resolved Big Bang.
At late times, a linear coasting cosmology $a(t) \propto t$ is obtained,
which is remarkably close to observation. The solution
consists of two sheets with opposite intrinsic chiralities, which are connected in a Euclidean pre-big bang era.

\end{abstract}


\maketitle

\subsection*{Introduction}

Quantum field theory and general relativity (GR) provide the basis of our present understanding 
of fundamental forces and matter. 
However, GR is a classical theory, and its incorporation into a consistent quantum theory poses fundamental challenges.
In particular, general arguments suggest  a ``foam-like''  quantum structure at the Planck scale $10^{-33}$ cm.  
One possible framework for such a description is provided by matrix models. Here we will focus on the IKKT model \cite{Ishibashi:1996xs}
which is singled out by maximal supersymmetry,  and is related to IIB string theory. In this approach, we should 
recover cosmological space-time as a solution, and the known physics should emerge from fluctuations on this background.

In this letter, we will give such a cosmological solution, which could be considered as near-realistic. 
It realizes a quantum
structure of space-time which is  exactly homogeneous and isotropic with $k=-1$,
with a finite density of microstates. 
A Big Bang (BB) arises through an appealing mechanism as in the $k=1$ 
solutions\footnote{Fuzzy cosmological solutions 
were also given in  \cite{Chaney:2015ktw}, but they are not fully homogeneous and isotropic in 3+1 dimensions.} 
\cite{Steinacker:2017vqw}, 
and the late-time evolution could be considered as near-realistic, corresponding to a 
coasting universe \cite{Melia:2011fj}. This has been discussed as a possible alternative to the 
$\Lambda$CDM model \cite{BenoitLevy:2011jt,Nielsen:2015pga}, 
however we refrain from claiming that the cosmology is fully realistic.
The background also leads to spin 2 fluctuations, which could be the basis for  emergent gravity, while
avoiding the issues (notably Wick rotation) which arose in a
previous semi-classical approach \cite{Klammer:2009ku}.
Quantum corrections are assumed to be small, which seems reasonable
for cosmological considerations.

We will consider solutions of the following IKKT - type matrix model 
\begin{align}
 S[Y] &= \frac 1{g^2}\Tr \Big([Y^a,Y^b][Y^{a'},Y^{b'}] \eta_{aa'} \eta_{bb'} \, - 2m^2 Y^a Y^b \eta_{ab}  \Big) \ . 
 \label{bosonic-action}
\end{align}
Here $\eta_{ab} = diag(-1,1,...,1)$ 
is interpreted as Minkowski metric of the target space $\R^{1,D-1}$. 
This captures the bosonic part 
 of the IKKT model \cite{Ishibashi:1996xs}, 
 supplemented by a mass term. 
This leads to the classical equations of motion 
\begin{align}
 \Box_Y Y^a + m^2 Y^a = 0 
 \label{eom-lorentzian-M}
\end{align}
where 
\begin{align}
  \Box_Y = \eta_{ab} [Y^a,[Y^b,.]]
  \label{Box-Y}
\end{align}
plays the role of the d'Alembertian.
These are also the effective eom for the IKKT model put forward in \cite{Kim:2012mw}, which arise
after taking into account an IR cutoff and integrating out the scale factor
in the matrix path integral 
\begin{align}
 Z = \int dY e^{i S[Y]}  
 \label{path-integral}
\end{align}
dropping the fermions for simplicity. 
The mass term  introduces a scale to the model, and it
arises naturally from an IR regularization 
as  discussed in \cite{Kim:2012mw}.

\subsection*{Euclidean fuzzy hyperboloids}

To define the fuzzy 4-hyperboloid $H^4_n$, let 
$\cM^{ab}$ be hermitian generators of $\mso(4,2) \cong \msu(2,2)$, which satisfy
\begin{align}
  [\cM_{ab},\cM_{cd}] &=i(\bar\eta_{ac}\cM_{bd} - \bar\eta_{ad}\cM_{bc} - \bar\eta_{bc}\cM_{ad} +\bar\eta_{bd}\cM_{ac}) \nn 
\end{align}
and $\bar\eta^{ab} = \diag(-1,1,1,1,1,-1)$ be the invariant metric.
We choose a particular type of (massless discrete series) 
positive-energy unitary irreps $\cH_{n}$ 
known as ``minireps'' or doubletons \cite{Mack:1975je,Gunaydin:1998sw,Govil:2013uta}, 
which have the remarkable property that they remain 
irreducible
under $SO(4,1) \subset SO(4,2)$. They have positive discrete spectrum 
\begin{align}
 {\rm spec}(\cM^{05}) = \{E_0, E_0+1, ... \}, \qquad E_0 = 1+\frac{n}2 
 \label{X0-discrete}
\end{align}
where the 
eigenspace with lowest eigenvalue of $\cM^{05}$ is an $n + 1$-dimensional irreducible
representation of either $SU (2)_L$ or $SU (2)_R$. Then the hermitian generators
\begin{align}
 X^a &:= r\cM^{a 5}, \qquad a = 0,...,4  \nn\\
   [X^a,X^b] &= - i r^2\cM^{ab}  =: i\Theta^{ab} 
\end{align} 
satisfy 
\begin{align} 
 \eta_{ab} X^a X^b &= X^i X^i - X^0 X^0 = - R^2 \one \ 
 \label{hyperboloid-constraint}
\end{align}
with $R^2 = r^2(n^2-4)$ \cite{Hasebe:2012mz}.
Since $X^0 = r \cM^{05} > 0$ has positive spectrum, this describes a one-sided 
Euclidean hyperboloid in $\R^{1,4}$, denoted as $H^4_n$. However 
the full semi-classical geometry underlying fuzzy  
$H^4_n$ is $\C P^{1,2}$ \cite{Hasebe:2012mz}. This is a coadjoint orbit $SU(2,2)/SU(1,2)\times U(1)$ , 
which is an
$SO(1,4)$-equivariant bundle over the 4-dimensional hyperboloid $H^4$ with fiber given by $S^2$.
Thus $X^a$ can be viewed as quantization\footnote{The appropriate $\msu(2,2)$ representations in the fuzzy case are obtained from \eq{Hopf-map}
if the $\psi$  satisfy a bosonic oscillator algebra, see e.g. \cite{Govil:2013uta}.} of the  Hopf map 
\begin{align} 
X^a \sim x^a = \obar\psi \Sigma^{a5} \psi: \ \C P^{1,2} \to H^4 \subset \R^{1,4}, 
\label{Hopf-map}
\end{align}
where $\psi\in \C^4$ transforms in the $(4)$ of $\msu(2,2)$ through $\Sigma^{ab}$.
We work mostly in this semi-classical limit, where
$H^4_n$ has the same local structure as the fuzzy 4-sphere $S^4_n$, with a
Poisson tensor $\cM^{\mu\nu} \sim  \theta^{\mu\nu}(x,\xi)$ transforming as a selfdual 2-form  under the local 
stabilizer\footnote{Note that the induced metric on the hyperboloid 
$H^4\subset \R^{1,4}$ is  Euclidean, despite of the $SO(4,1)$ isometry;
this is obvious at the point $x=(R,0,0,0,0)$.}  $SO(4)_x$ 
of any point $x \in H^4$. This
realizes the local  fiber $\xi \in S^2$, which  in the fuzzy case is  fuzzy $S^2_n$. 
Then the averaging over this $S^2$
 can be achieved using the same local formula  \eq{average-H} as for $S^4_N$, which will be important below.

In particular, $H^4_n$ has a finite density of microstates, since
the number of states in $\cH_n$ between two given $X^0$-eigenvalues is finite.
Note that $n$ is not required to be large here, in contrast to fuzzy $S^4_N$.

\subsection*{Cosmology from squashed fuzzy hyperboloid $H^4_n$}

It was shown in \cite{Steinacker:2017vqw} that in the presence of 
$SO(4,1)$-breaking mass terms, the matrix model has
solutions based on $H^4_n$ which describe
fuzzy FRW space-times  with Minkowski signature. 
Here we insist on invariant mass terms, 
and look for solutions  of the matrix model  \eq{bosonic-action}
given by rescaled generators $Y^\mu, \ \mu\in\{0,2,3,4\}$
\begin{align}
   Y^\mu &=  X^\mu , \quad \mbox{for} \ \  \mu = 0,2,3,4 .
 \label{H4-ansatz}
\end{align}
Note that the generator $X^1$ is dropped. 
The indices $i$ indicate Euclidean directions, and $0$ is  the time-like direction.
Geometrically (i.e. in the semi-classical limit), 
this corresponds to a brane $\cM$ which is 
embedded in $\R^{1,3}$ via a projection $\Pi$ along $x^1$,
\begin{align*}
 Y^\mu \sim y^\mu: \  \C P^{1,2} \to  H^4 \ \stackrel{\Pi}{\longrightarrow} \ \R^{1,3} \ .
\end{align*}
Thus $\cM = \cM^+ \cup \cM^-$ consists of two sheets which are connected at the boundary.
 The $\mso(2,4)$ algebra  gives
\begin{align}
  [Y^\mu,[Y^\mu,Y^\nu]] &= -i r^2 [Y^\mu,\cM^{\mu\nu}]  \qquad\mbox{(no sum)}  \nn\\
    &=  r^2 \left\{\begin{array}{ll}
               - Y^\nu, & \nu\neq \mu \neq 0 \\
                 Y^\nu, & \nu\neq \mu = 0 \\
               0, & \nu = \mu
              \end{array}\right. \ 
\end{align}
so that
\begin{align}
 \Box_Y Y^\mu = [Y^\nu,[Y_\nu,Y^\mu]] &= -3 r^2 Y^\mu \ .
\end{align}
Therefore the $Y^\mu$ for $\mu = 0,2,3,4$ are a solution of \eq{eom-lorentzian-M}
with $m^2 =  3 r^2$. This is the solution of interest here. 
Note that $m^2>0$ suggests stability of the background, which however needs to be 
studied through a complete fluctuation analysis elsewhere.

\begin{figure}[ht]
\vspace{0.5cm}
\begin{center}
\includegraphics[scale=0.5]{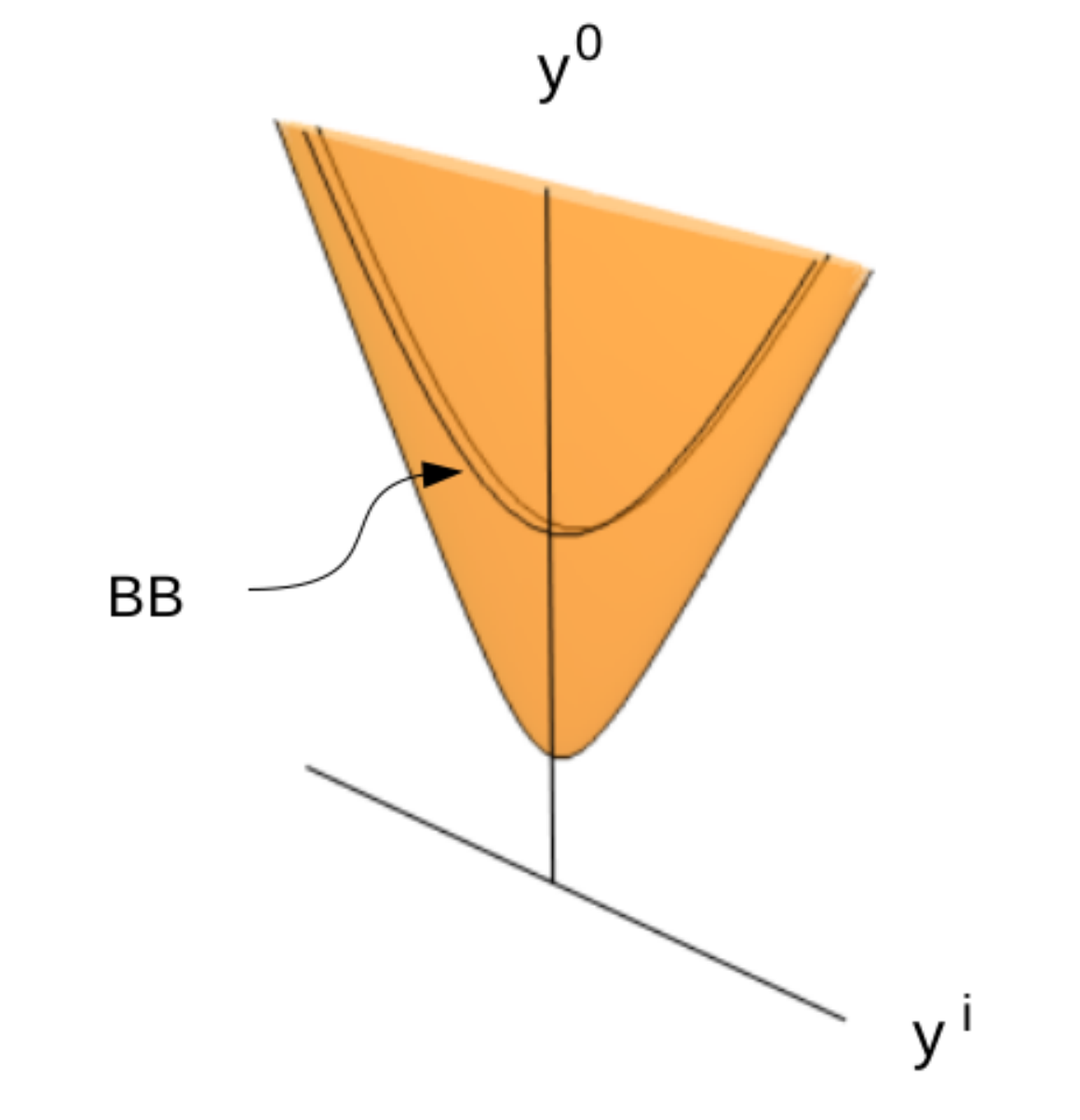}
\end{center}
\vspace{-0.2cm}
\caption{Two-sheeted universe $\cM$ in the $(y^0,y^i)$ plane, with Big Bang (BB). }
\vspace{-0.2cm}
\label{fig:sheets}
\end{figure}

These $Y^\mu$ transform as vectors of $SO(1,3)$, implemented by  the adjoint i.e. by gauge transformations.
Hence the solution admits a global $SO(1,3)$ symmetry. 
In the semi-classical limit, this defines a foliation 
\begin{align}
\cM \ = \  \Pi(H^4) \ = \ \cM^+ \cup_{t_0} \cM^- \ =  H^3_{t_0} \ \cup \  2\bigcup\limits_{t > t_0} H^3_{t} \nn
\end{align}
into $SO(3,1)$-invariant space-like 3-hyperboloids $H^3_t$ (one for each sheet except for $t_0$), whose
effective metric depends on the time parameter $t$. 
The two sheets $\cM^+ \cup_{t_0} \cM^-$ are connected at $t=t_0$, indicated by $\cup_{t_0}$,
as indicated in figure \ref{fig:sheets}. We obtain a double-covered
FRW space-time with hyperbolic ($k=-1$) spatial geometries.

%

\paragraph*{Induced metric.}

The induced metric on $\cM\subset \R^{1,3}$ is
\begin{align}
 g_{\mu\nu} = (-1,1,1,1) = \eta_{\mu\nu} , \qquad \mu,\nu=0,2,3,4  \ .
\end{align}
We can write this as a $SO(3,1)$-invariant FRW metric with $k=-1$,
decomposing $\cM$ into 3-hyperboloids $H_t$
\begin{align}
\begin{pmatrix}
 y^0 \\ y^2 \\ y^3 \\ y^4
\end{pmatrix}
 = t\, \begin{pmatrix}
  \cosh(\chi) \\
   \sinh(\chi)\begin{pmatrix}
                                               \sin(\theta)\cos(\varphi)\\
                                               \sin(\theta)\sin(\varphi)\\
                                               \cos(\varphi)
                                              \end{pmatrix} 
\end{pmatrix}                                             
 \label{local-hyperbolic-coords}
\end{align}
for $t \geq 0$.
This gives the Milne metric:
\begin{align}
 ds_g^2 &= -d t^2 + t^2\, d\Sigma^2 = -dy_0^2 + dy_2^2+ dy_3^2+dy_4^2, \nn\\
 d\Sigma^2 &= d\chi^2 + \sinh(\chi)^2 d\Omega^2 \ .
 \label{milne-metric}
\end{align}
Here $d\Sigma^2$ is the $SO(3,1)$-invariant metric on a space-like $H^3$ with $k=-1$,
and  $d\Omega^2$ is the $SO(3)$-invariant metric on the unit sphere $S^2$.

The induced metric can be viewed as closed-string metric in target space.
However the fluctuations on the brane are governed by a different effective metric,
which can be viewed as open string metric \cite{Seiberg:1999vs}:

\paragraph*{Effective metric.}

The effective metric encoded in $\Box_Y$ is given by \cite{Steinacker:2017vqw,Steinacker:2016vgf,Steinacker:2010rh}
\begin{align}
  G^{\mu\nu} &= \a\, \g^{\mu\nu} \ , \qquad \a = \sqrt{\frac{|\theta^{\mu\nu}|}{|\g^{\mu\nu}|}} \ , \nn\\
 \g^{\mu\nu} &= g_{\mu'\nu'}[\theta^{\mu'\mu}\theta^{\nu'\nu}]_{S^2} 
   \label{eff-metric-G}
\end{align}
where $[]_{S^2}$ indicates averaging over the internal $S^2$. To understand this, it is useful to
consider first the unprojected $H^4$. Then this averaging is given by 
\begin{align}
  \left[  \theta^{ab}_{\bf x}  \theta^{cd}_{\bf x}\right]_{S^2} 
  &=\frac{1}{12} \D^4  (P^{ac}_H P^{bd}_H - P^{bc}_H P^{ad}_H \pm \varepsilon^{abcde} \frac 1{R} x^e) ,
\label{average-H}
\end{align}
where 
\begin{align}
 P^{ac}_H = \eta^{ac} + \frac 1{R^2} x^a x^c, \qquad  \eta_{ab} x^a x^b = -R^2  \nn
\end{align}
is the $SO(4,1)$-invariant projector on the tangent space of $H^4\subset \R^{1,4}$,
with Cartesian coordinates $x^a$ as above. Note that
\eq{average-H} is the unique $SO(4,1)$- invariant  tensor incorporating the appropriate antisymmetry and the
selfduality of  $ \theta^{ab}_{\bf x}$
(recall that $H^4$ is Euclidean w.r.t. $ \eta_{ab}$).

Now consider the projected case. The projection $\Pi$ to $\R^{1,3}$ in \eq{average-H} is 
achieved simply by  dropping the $x^1$ components; in particular,
$\theta^{\mu\nu}_{\bf y} = \{y^\mu,y^\nu\}$ 
is obtained from $\theta^{ab}_{\bf x}$ simply by dropping these indices,  in the 
$y^\mu$ coordinates.
We shall evaluate this at a reference point 
\begin{align}
 x=(x_0,x_1,0,0,0) = R(\cosh(\eta),\sinh(\eta),0,0,0) \ \in H^4 , \nn 
\end{align}
which is projected to
\begin{align}
 y = \Pi(x) = (y_0,0,0,0) = R(\cosh(\eta),0,0,0) \ \in \R^{1,3} . \nn  
\end{align}
We need 
\begin{align}
 P^{00}_H &= -1 + \frac 1{R^2} y^0 y^0 = -1+\cosh^2(\eta), \nn\\ 
 P^{ii}_H &= 1  \nn \ 
\end{align}
so that
\begin{align}
g_{\mu\nu} P_H^{\mu\nu} &= 4 - \frac{1}{R^2} y_0^2    \
= 4 -\cosh^2(\eta)   \nn
\end{align}
and
\begin{align}
\g^{\mu\nu}  &= \frac{1}{12} \D^4 \Big((4 -\cosh^2(\eta)) P^{\mu\nu}_H - g_{\rho\s} P^{\rho\mu}_H P^{\s\nu}_H \Big)  \ . \nn
\end{align}
This gives
\begin{align}
 \g^{ii} 
         &=  \frac{ \D^4}{12} (3 -\cosh^2(\eta)) =:   \frac{\D^4}{4}  c(\eta) \nn\\
         &=  \frac{ \D^4}{4} (1- \frac 13 \frac{y_0^2}{R^2}) =: \frac{\D^4}{4} c(y_0) \nn\\
\g^{00} 
         &=  \frac{\D^4}{4}  (\cosh^2(\eta)-1 ) =:  \frac{\D^4}{4} c_0(\eta)    \nn\\
         &= \frac{\D^4}{4}(\frac{y_0^2}{R^2}-1) =: \frac{\D^4}{4} c_0(y_0)  
\end{align}
where the  time parameters $y_0$ or $\eta$ will be used appropriately.
Therefore
\begin{align}
 \g^{\mu\nu} =  \frac{\D^4}{4} (c_0(\eta), c(\eta), c(\eta), c(\eta)) \ .
\end{align}
While  the first component is non-negative, there is a signature change at $c(\eta)=0$ i.e.
\begin{align}
 \cosh^2(\eta_0)=3
 \label{eta-0}
\end{align}
which marks the Big Bang. The effective metric is Euclidean for $\eta<\eta_0$, 
and Minkowskian for $\eta > \eta_0$.
In contrast, the induced metric always has Minkowski signature.

Now consider the conformal factor $\a$  \eq{eff-metric-G}. Since
the (Kirillov-Kostant) symplectic form $\omega$ on $\C P^{1,2}$ is $SO(4,2)$-invariant and 
using $y_0 = R\cosh(\eta)$ and
$dy_0 = R\sinh\eta d\eta =  \sinh(\eta)\, d\tau$ where $\tau= R\eta$,
the  $SO(4,1)$-invariant volume form on $H^4$ can be written as
\begin{align*}
 \omega^2|_{H^4} = \frac 4{\D^4} d\tau dy_2...dy_4 &= \frac 4{\D^4} \sinh(\eta)^{-1}   dy_0 ... dy_4   \nn\\
  &= \sqrt{|\theta^{-1}_{\mu\nu}|} dy_0 ... dy_4 \ .
\end{align*}
Therefore 
\begin{align*}
 \sqrt{|\theta^{\mu\nu}|} =  \frac{\D^4}4  \sinh(\eta)
  = \frac{\D^4}4 \sqrt{c_0(\eta)}  \ .
\end{align*}
Hence the conformal factor is
\begin{align*}
 \a  &=  \sqrt{\frac{|\theta^{\mu\nu}|}{|\g^{\mu\nu}|}}
  =  \frac{4}{\D^4} |c(\eta)|^{-\frac{3}{2}} 
\end{align*}
assuming $\eta > \eta_0$.
The effective metric is obtained as
 \begin{align}
  G_{\mu\nu} 
  &=  \Big(\frac{|c(\eta)|^{\frac{3}{2}}}{c_0(\eta)},-|c(\eta)|^{\frac{1}{2}} ,
    -|c(\eta)|^{\frac{1}{2}},-|c(\eta)|^{\frac{1}{2}}\Big)  \nn\\
  &\sim y_0 \big(\frac 13,-1,-1,-1\big),  \qquad \frac{y_0}R \gg 1 \ . 
 \label{eff-metric-hyperbel}
\end{align}
However, this form is only valid in the  local Cartesian coordinates $y^\mu$ at the reference point 
$p\in H^3_{\tau} \subset\cM$.

\paragraph*{Scale factor.}

Now we rewrite the above metric from local Cartesian coordinates $y^\mu$
into FRW coordinates $(\tau,\chi,\theta,\varphi)$ \eq{local-hyperbolic-coords}.
{\em At the reference point} $p$ with $\chi=0$, the $SO(1,3)$-invariant metric on $H^3_\tau$
has the form
\begin{align}
ds^2|_{H^3} = \sum_i dy_i^2 = y_0^2 d\Sigma^2
\end{align}
where $d\Sigma^2$ is the length element on a spatial standard 3-hyperboloid $H^3$.
Therefore at the reference point
\begin{align}
 ds^2_G|_p &= G_{\mu\nu} dy^\mu dy^\nu 
 = \frac{|c(y_0)|^{\frac{3}{2}}}{c_0(y_0)} dy_0^2 - |c(y_0)|^{\frac{1}{2}} y_0^2\, d\Sigma^2 \ .
 \nn   
\end{align}
 Thus the FRW form of the effective metric is obtained as
 \begin{align}
 ds^2_G &= d t^2 - a^2(t)d\Sigma^2 \ 
\end{align}
where the scale parameter $a(t)$ is determined by 
\begin{align*}
 \frac{d y_0}{d t} &= \frac{c_0(y_0)^{\frac{1}{2}}} {|c(y_0)|^{\frac{3}{4}}} \ ,
 \qquad a(t) = |c(y_0)|^{\frac{1}{2}} y_0^2 \ .
\end{align*}
A closed form for the desired scale function $a(t)$ is obtained only for early and late times.
For late times $y_0 \to \infty$, the metric simplifies as
\begin{align*}
 ds_G^2 &= \frac 13 y_0 d y_0^2 - y_0^3\, d\Sigma^2 = d t^2 - a(t)^2 d\Sigma^2 \ 
\end{align*}
dropping $R$ for simplicity. 
We thus read off
\begin{align*}
 \frac 1{\sqrt{3}} \sqrt{y_0}\, d y_0 
 = dt, \qquad t= \frac 2{3\sqrt{3}}\, y_0^{3/2} 
\end{align*}
and therefore 
\begin{align}
 a(t)  = y_0^{3/2} = \frac{3\sqrt{3}}{2} t \ .
 \label{a-late-coast}
\end{align}
This describes a linear coasting universe \cite{Melia:2011fj,Melia:2014aja}, 
which provides a remarkably good fit with observation \cite{Nielsen:2015pga,Melia:2011fj}.
It has been considered \cite{Nielsen:2015pga} and disputed \cite{Haridasu:2017lma}
as a possible alternative to the  $\Lambda$CDM model. 
While such a cosmology seems artificial within GR, the present framework provides a
good theoretical basis, and should motivate a careful re-assessment of this scenario.
In particular, the age of the Universe is correctly reproduced as $\approx 13.9 \times 10^9$ years
from the present Hubble parameter,
\begin{align*}
 H = \frac{\dot a}{a} = t^{-1} \ .
\end{align*}
If  $\dot a$ were 1, we would recover the  Milne universe \eq{milne-metric}. However \eq{a-late-coast}
gives a larger spatial curvature radius,  which is large compared with the scale of the visible universe. 
This modification might have some effect e.g. on the CMB, and could have an impact on the detailed assessment of the model.
To address this in a reliable way however requires a more detailed analysis of the physics on the
present background. 

Now consider the Big Bang singularity.
Shortly after the Big Bang $\eta \gtrsim \eta_0$ \eq{eta-0}, we can write 
\begin{align}
 dt &=  \frac{|c(y_0)|^{\frac{3}{4}}}{c_0(y_0)^{\frac{1}{2}}}  d y_0
 \propto \big(\frac{y_0}R -\sqrt{3}\big)^{\frac{3}{4}} dy_0
 \ \propto\ R  d\big(\frac{y_0}R  -\sqrt{3}\big)^{\frac{7}{4}}, \nn\\
 t & \propto  \big(\frac{y_0}R -\sqrt{3}\big)^{\frac{7}{4}}
 \label{t-y0-BB}
 \end{align}
and therefore 
\begin{align*}
 a(t) \propto \, c(t)^{\frac{1}{4}}\ \propto \  t^{1/7}
\end{align*}
dropping some constants.
Hence we recover the same initial  $a(t)\sim t^{1/7}$ expansion as in \cite{Steinacker:2017vqw}.
The physical implications of this singular  but non-exponential initial expansion remain to be understood.
Recall that the BB arises only through the effective metric as in \cite{Steinacker:2017vqw}, 
while there is no singular expansion and no signature change in the induced  metric.
Together with a Euclidean pre-Big Bang era, this should have interesting implications
on the early cosmology including the CMB, and  resolve
the horizon problem even in the absence of standard inflation.
Note that all propagating modes on $\cM$, including the gravitational modes, 
are governed by the effective metric, as discussed in the next section.

\section*{Further perspectives}
\label{sec:discussion}

The effective physics on this background $\cM$ should be elaborated through a careful fluctuation analysis
of the bosonic and fermionic modes. In the matrix model, 
these fluctuations can be understood in terms of  the algebra of functions  $End(\cH_n) \sim Fun(\C P^{1,2})$ 
on $H^4_n$, which reduces in the semi-classical limit to the algebra of functions on $\C P^{1,2}$. Since the latter is 
an equivariant (``twisted'') bundle over $H^4$, the harmonics on the fiber $S^2$ lead to higher spin modes on $H^4$, 
in contrast to  Kaluza-Klein modes which arise on ordinary compactifications.
Explicitly,  functions $\Phi \in End(\cH_n)$ can be expanded in the form
\begin{align}
 \Phi = \phi(x) + \phi_{ab}(x) \cM^{ab} + ...  \ .
 \label{phi-modes}
\end{align}
As explained in \cite{Sperling:2017gmy,Steinacker:2016vgf}, this amounts to a decomposition 
into higher spin modes,
whose propagation is governed by $\Box_Y$ \eq{Box-Y}, hence by the effective metric $G_{\mu\nu}$.
The maximal spin in \eq{phi-modes} is  $n$, determined by the maximal mode on the internal fuzzy sphere $S^2_n$. In particular, 
the tangential fluctuations $Y^\mu + \cA^\mu$ include the following modes
\begin{align*}
 \cA^\mu = \theta^{\mu\nu} h_{\nu\rho}(x) P^\rho
\end{align*}
where $P^\mu \in \mso(4,1)$ is the local generator of translations, cf. \cite{Steinacker:2016vgf}.
The symmetric part of $h_{\mu\nu}$ could naturally play the role of the spin 2 graviton. 
Whether or not this leads to physically viable gravity is a non-trivial question,
which should be examined in detail elsewhere, cf. \cite{Sperling:2017gmy,Steinacker:2016vgf}. Here we only note that  
the issues encountered on $S^4_N$ may not arise here since $n$ can be small.
If this emergent gravity works out correctly below the cosmological scale, 
this could provide an elegant solution of the cosmological 
constant or dark energy problem.

Once matter and fields on $\cM$ are  taken into account, the energy density near the BB 
would formally diverge in the semi-classical picture. 
However, the inherent discreteness  of $y_0$ \eq{X0-discrete} leads via \eq{t-y0-BB}
to a natural discretization of $t$,  hence to a UV cutoff. 
At this point any continuous evolution equation will break down, 
and must be replaced by some (yet to be determined) boundary condition relating the Euclidean and Minkowski regimes.
Nevertheless, the high energy density near the BB will certainly have a significant effect on the geometry,
which  remains to be clarified.

Another interesting aspect of the present solution is the two-sheeted structure of $\cM$. 
Since the bundle of Poisson tensors
$\theta^{\mu\nu}$ on $H^4_n$ is self-dual on one sheet and anti-selfdual on the other,
this could provide the seed for a chiral gauge theory, e.g. through left-right symmetric 
models.
This may also provide a dynamical justification for the present solution 
over e.g. a Euclidean $H^4$ solution where all five $X^a$ are embedded, since  branes with opposite chirality 
tend to form a bound state, cf. \cite{Chepelev:1997av,Steinacker:2015dra}.
The additional structure required for particle physics might  arise 
from the remaining 6 dimensions in the matrix model, perhaps along the lines of 
\cite{Steinacker:2014lma,Chatzistavrakidis:2011gs}. 
It is thus quite conceivable that all fundamental interactions could arise from a 
suitable refinement of the present background.

\paragraph*{Acknowledgements.}

I would like to thank in particular Hikaru Kawai, Jun Nishimura and Marcus Sperling for very useful discussions.
This work was fully funded by the Austrian Science Fund (FWF) grant P28590. 
The Action MP1405 QSPACE from the European Cooperation in Science and Technology (COST) also
provided support in the context of this work.

\appendix

\end{document}